\documentclass[9pt,conference]{IEEEtran}
\usepackage{waspaa25}

\ifCLASSINFOpdf
\else
   \usepackage[dvips]{graphicx}
\fi
\usepackage{url}
\usepackage{amsmath}
\usepackage{amssymb}
\usepackage{adjustbox}
\usepackage{graphicx}
\usepackage[utf8]{inputenc}
\usepackage[nolist,nohyperlinks]{acronym}

\makeatletter
\AtBeginDocument
 {
   \def\ltx@label#1{\cref@label{#1}}
   \def\label@in@display@noarg#1{\cref@old@label@in@display{#1}}
\def\label@in@mmeasure@noarg#1{%
    \begingroup%
      \measuring@false%
      \cref@old@label@in@display{#1}
    \endgroup}%
 } %
\makeatother

\begin{acronym}
    \acro{DDSP}{Differentiable Digital Signal Processing}
\acro{NMOS}{Noise \acs{MOS}}
\acro{SMOS}{Speech \acs{MOS}}
\acro{GMOS}{Global \acs{MOS}}
\acro{SCNR}{Single-channel \ac{NR}}
\acro{NR}{Noise Reduction}
\acro{TED}{Technology Engineering for Devices}
\acro{NN}{Neural Network}
\acro{MIPS}{Million Instructions Per Second}
\acro{SE}{Speech Enhancement}
\acro{DL}{Deep Learning}
\acro{PMWF}{Parameterized Multi-channel Wiener Filter}
\acro{SI-SDR}{Scale-Invariant Signal-to-Distortion Ratio}
\acro{MAC}{Multiply Accumulate Operation}
\acro{GRU}{Gated Recurrent Unit}
\acro{PCM}{Phase-constrained Magnitude}
\acro{MCSE}{Multi-channel Speech Enhancement}

\acro{3GPP}{3rd Generation Partnership Project}
\acro{a-SLAM}[aSLAM]{Acoustic \aclu{SLAM}}
\acro{aSLAM}{Acoustic \aclu{SLAM}}
\acro{A-SNR}{A-weighted \acs{SNR}}
\acro{AAC}{Advanced Audio Coding\acroextra{. A lossy codec used for digital audio.}}
\acro{AAD}{Auditory Attention Detection}
\acro{AAS}{American Auditory Soc.}
\acro{AASP}{Audio and Acoustic Signal Processing}
\acro{ABC}{Analytical with or without Bias Compensation}
\acro{ABR}{Auditory-Brainstem Response}
\acro{ACAWD}{Archivable Core Actual-Word Database}
\acro{ACB}{Adaptive Codebook}
\acro{ACC}{Accuracy}
\acro{ACE}{Acoustic Characterization of Environments\acroextra{. A noisy reverberant speech corpus and IEEE challenge run by the SAP group at Imperial College}}
\acro{ACELP}{Algebraic Code-Excited Linear Prediction}
\acro{ACF}{Autocorrelation Function}
\acro{ACK}{Acknowledgement}
\acro{ACL}{Access Control List}
\acro{ACR}{Absolute Category Rating}
\acro{AD}{Audio Diarization}
\acro{ADC}{Analogue-to-Digital Converter}
\acro{ADM}{Adaptive Differential Microphone}
\acro{ADPCM}{Adaptive Differential Pulse Code Modulation}
\acro{ADSL}{Asymmetric Digital Subscriber Line}
\acro{AE}{Almost Everywhere}
\acro{AES}{Audio Engineering Society}
\acro{AES2}[AES]{Advanced Encryption Standard}
\acro{AGC}{Automatic Gain Control}
\acro{AH}{Amplitude Histogram}
\acro{AI}{Articulation Index}
\acro{AI2}[AI]{Artificial Intelligence}
\acro{AI3}[AI]{Audio Inpainting}
\acro{AIC}{Akaike Information Criterion}
\acro{AIFF}{Audio Interchange File Format}
\acro{AIR}{Acoustic Impulse Response}
\acro{AIR2}[AIR]{Aachen Impulse Response}
\acro{AIRD}{Aachen Impulse Response Database}
\acro{ALC}{Automatic Level Control}
\acro{ALCons}{Articulation Loss of Consonants}
\acro{AM}{Amplitude Modulation}
\acro{AMDF}{Average Magnitude Difference Function\acroextra{. A function with similar properties to the cross- or autocorrelation but that requires no multiplication to evaluate.}}
\acro{AMR}{Adaptive Multi-Rate}
\acro{AMR-NB}{Adaptive Multi-Rate Narrow Band}
\acro{AMR-WB}{Adaptive Multi-Rate Wide Band}
\acro{AMS}{Amplitude Modulation Spectrogram}
\acro{ANC}{Adaptive Noise Canceller}
\acro{ANS}{Autocorrelation-Based Noise Subtraction}
\acro{ANSI}{American National Standards Institute}
\acro{ANU}{Australian National University}
\acro{APLAWD}{Archivable Priority List Actual-Word Database}
\acro{AR}{Autoregressive}
\acro{ARD}{Arbeitsgemeinschaft der \"{o}ffentlich-rechtlichen Rundfunkanstalten der Bundesrepublik Deutschland\acroextra{. `Consortium (``Working group'') of the public-law broadcasting institutions of the Federal Republic of Germany'}}
\acro{ARMA}{Autoregressive Moving Average}
\acro{AS}{Audio Segmentation}
\acro{AS2}[AS]{Almost Surely}
\acro{ASA}{Acoustic Scene Analysis}
\acro{ASIO}{Audio Stream Input/Output\acroextra{. A computer soundcard protocol with low latency developed by Steinberg.}}
\acro{ASK}{Amplitude Shift Keying}
\acro{ASLP}{Audio, Speech, and Language Processing}
\acro{ASM}{Acoustic Scene Mapping}
\acro{ASR}{Automatic Speech Recognition}
\acro{ASS}{Approximate Spectrum Substitution}
\acro{AST}{Acoustic Source Tracking}
\acro{AST2}[AST]{Asymmetric Sampling in Time}
\acro{ATF}{Acoustic Transfer Function\acroextra{. The Fourier Transform of the \acs{RIR}.}}
\acro{ATLM}{Acoustic Tokenization and Language Modelling}
\acro{ATR}{Advanced Telecommunications Research Institute International\acroextra{, Kyoto, Japan}}
\acro{AUC}{Area under the curve}
\acro{AURORA}{Aurora Experimental Framework for the Evaluation of the Performance of Speech Recognition Systems under Noisy Conditions}
\acro{AUV}{Autonomous Underwater Vehicle}
\acro{AV}{Audio-Visual}
\acro{AWGN}{Additive White Gaussian Noise}
\acro{AWS}{Approximate Waveform Substitution}
\acro{BASIE}{Bayesian Adaptive Speech Intelligibility Estimation}
\acro{BCE}{Blind Channel Estimation}
\acro{BCL}{Bekesy Comfortable Loudness}
\acro{BCR}{Block-Coordinate Relaxation}
\acro{BEM}{Boundary Element Method}
\acro{BER}{Bit Error Rate}
\acro{BIBO}{Bounded-Input Bounded-Output}
\acro{BIC}{Bayesian Information Criterion}
\acro{Bk}{Berksons\acroextra{. A unit for measuring intelligibility.}}
\acro{BK}[B\&K]{Br{\"u}el and Kj{\ae}r}
\acro{BLSTM}{Bidirectional \acs{LSTM}}
\acro{BM}{Blocking Matrix}
\acro{BO-SCPHD}{Bearing-only \acs{SC-PHD}}
\acro{BO-SLAM}{Bearing-only \acs{SLAM}}
\acro{BOT}{Bearing-only tracking}
\acro{BP}{Basis Pursuit}
\acro{BPCC}{Basis Pursuit with Clipping Constraints}
\acro{BPDN}{Basis Pursuit Denoising}
\acro{BPM}{Beats Per Minute}
\acro{BPSK}{Binary Phase Shift Keying}
\acro{BR}{Barrodale and Roberts' (algorithm)}
\acro{BRI}{Basic Rate Index}
\acro{BSD}{Bark Spectral Distortion}
\acro{BSI}{Blind System Identification}
\acro{BSIM}{Binaural Speech Intelligbility Model}
\acro{BSS}{Blind Source Separation}
\acro{BSTOI}{Binaural \acs{STOI}}
\acro{BW}{Bandwidth}
\acro{BZ}{Back-to-Zero}
\acro{C4DM}{Centre for Digital Music}
\acro{C50}[$C_\textrm{50}$]{Clarity Index}
\acro{C-GFB}{Combination Gas-fired Boiler}
\acro{CART}{Classification and Regression Tree}
\acro{CASA}{Computational Auditory Scene Analysis}
\acro{CBR}{Constant Bit Rate}
\acro{CCC}{Cross-Correlation Coefficient}
\acro{CCCC}{DARPA CSR Corpus Coordinating Committee}
\acro{CCD}{Charge-Coupled Device}
\acro{CCI}{Call Clarity Index}
\acro{CCITT}{Consultative Committee for International Telephony and Telegraphy}
\acro{CCM}{Contralateral Competing Message}
\acro{CCR}{Comparison Category Rating}
\acro{CDB}{Constant Directivity Beamformer}
\acro{CDMA}{Code Division Multiple Access}
\acro{CELP}{Code-excited Linear Prediction}
\acro{CHIEF}{Combined Helmholtz Integral Equation Formulation}
\acro{CIT}{Constrained Initial Taps}
\acro{CL}{Clipping Level}
\acro{CLEAR}{Centre for Law Enforcement Audio Research}
\acro{CLID}{Cluster Identification Test}
\acro{CLT}{Central Limit Theorem}
\acro{CMA}{Constant Modulus Algorithm}
\acro{CMASI}{Coherence Modulated Acoustic Speckle Interferometry}
\acro{CMB}{Cosmic Microwave Background}
\acro{CNC}{Consonant-Nucleus-Consonant}
\acro{CNG}{Comfort Noise Generation}
\acro{CNN}{Convolutional Neural Network}
\acrodefplural{CNN}[CNNs]{Convolutional Neural Networks}
\acro{CODEC}{Coder-Decoder}
\acro{CPE}{Customer Premises Equipment}
\acro{CPHD}{Cardinalized \acs{PHD}}
\acro{CRACD}{Codec-robust Automatic Clipping Detector}
\acro{CRC}{Cyclic Redundancy Check}
\acro{CRNN}{Convolutional Recurrent Neural Network}
\acro{CS}{Channel Shortening}
\acro{CS2}[CS]{Compressive Sensing}
\acro{CSP}{Communications and Signal Processing}
\acro{CSR-WSJ}{Continuous Speech Recognition Wall Street Journal Phase 1\acroextra{ database}}
\acro{CST}{Connected Speech Test\acroextra{ speech corpus}}
\acro{CT}{Conversation Test}
\acro{CTTN}{Comparative Tolerance to Noise}
\acro{CV}{Coefficient of Variation}
\acro{CVNN}{Complex-Valued Neural Networks}
\acrodefplural{CV}{Coefficients of Variation}
\acro{CV2}[CV]{Constant Velocity}
\acro{CVC}{Consonant-Vowel-Consonant}
\acro{CW}{Continuous Wave}
\acro{CWM}{Centre-Weighted Median}
\acro{CWMY}{Centre-Weighted Myriad}
\acro{CWT}{Continuous Wavelet Transform}
\acro{DAC}{Digital-to-Analogue Converter}
\acro{DAM}{Diagnostic Acceptability Measure}
\acro{DAQ}{Data Acquisition}
\acro{DARPA}{Defense Advanced Research Projects Agency\acroextra{ of the United States Dept. of Defense}}
\acro{DARPA-RMD}{\acs{DARPA} 1000-Word Resource Management Database\acroextra{ for Continuous Speech Recognition}}
\acro{DAW}{Digital Audio Workstation}
\acro{dB}{Decibel}
\acro{dBFS}{\acs{dB} Full Scale}
\acro{DBN}{Deep Belief Network}
\acro{DBSTOI}{Deterministic \acs{BSTOI}}
\acro{DC}{Direct Current}
\acro{DCME}{Digital Circuit Multiplexing Equipment}
\acro{DCR}{Degradation Category Rating}
\acro{DCT}{Discrete Cosine Transform}
\acro{DDR}{Direct-to-diffuse ratio}
\acro{DDR3}{Double Data Rate Type Three}
\acro{DECT}{Digital European Cordless Telecommunication}
\acro{DeLILAH}{Detection of Clipping using Least Squares Residuals and Iterated Logarithm Amplitude Histogram}
\acro{DENBE}{\acs{DRR} Estimation using a Null-Steered Beamformer}
\acro{DET}{Detection Error Trade-off}
\acro{Dev}{Development\acroextra{ dataset of the \acs{ACE} Challenge}}
\acro{DFT}{Discrete Fourier Transform}
\acro{DI}{Directivity Index}
\acro{DI-NN}{Dual-Input Neural Network}
\acro{DirectX}{\acroextra{A programming interface developed by Microsoft for handling tasks related to multimedia.}}
\acro{DIRHA}{Distant-speech Interaction for Robust Home Applications\acroextra{ multi-microphone multi-language acoustic speech corpus}}
\acro{DMA}{Distributed Microphone Array}
\acro{DMA2}{Differential Microphone Array}
\acro{DMT}{Discrete Multi-Tone}
\acro{DMV}{Dynamically Managed Voice\acroextra{ system}}
\acro{DNN}{Deep Neural Network}
\acro{DNR}{Dynamic Noise Reduction}
\acro{DoA}{Direction-of-Arrival}
\acrodefplural{DoA}[DoAs]{Directions-of-Arrival}
\acro{DOA}{Direction-of-Arrival}
\acrodefplural{DOA}[DOAs]{Directions-of-Arrival}
\acro{DP}{Dynamic Programming}
\acro{DPCM}{Differential Pulse Code Modulation}
\acro{DPD}{Direct-Path Dominance}
\acro{DPD-MUSIC}{Direct-Path Dominance Multiple Signal Classification}
\acro{DR}{Douglas-Rachford}
\acro{DR2}{Dynamic Range}
\acro{DRM}{Diagnostic Rhyme Test}
\acro{DRR}{Direct-to-Reverberant Ratio}
\acro{DRNN}{Deep Recurrent Neural Net}
\acro{DRT}{Diagnostic Rhyme Test}
\acro{DSB}{Delay-and-Sum Beamformer}
\acro{DSOBM}{Deterministic \acs{SOBM}}
\acro{DSP}{Digital Signal Processing}
\acro{DSPS}{Double Sides Periodic Substitution}
\acro{DSR}{Distributed Speech Recognition}
\acro{DSWS}{Double Sides Waveform Substitution}
\acro{DTMF}{Dual Tone Multi-Frequency}
\acro{DTX}{Discontinued Transmission}
\acro{DWT}{Discrete Wavelet Transform}
\acro{EARS}{Embodied Audition for RobotS}
\acro{EBF}{Eigen-beamformer}
\acro{EBU}{European Broadcasting Union}
\acro{EC}{Echo Canceller}
\acro{EDC}{Energy Decay Curve}
\acro{EDR}{Energy Decay Relief}
\acro{EDF}{Energy Decay Function}
\acro{EEG}{Electroencephalography}
\acro{EER}{Equal Error Rate}
\acro{EFICA}{Efficient Fast Independent Component Analysis}
\acro{EF-NN}{Early Fusion Neural Network}
\acro{EIR}{Equalized Impulse Response}
\acro{EKF}{Extended Kalman Filter}
\acro{EKF-SLAM}[EKF-SLAM]{\acs{EKF} \acs{SLAM}}
\acro{EKF-SLAM2}[EKF-SLAM]{Extended Kalman Filter \acs{SLAM}}
\acro{EL}{Echo Loss}
\acro{ELF}{Extremely Low Frequency}
\acro{ELRA}{European Languages Research Association}
\acro{EM}{Estimation-Maximization\acroextra{. An iterative technique to solve certain optimization problems.}}
\acro{EMIB}{Eigenmike Microphone Interface Box}
\acro{ENF}{Electrical Network Frequency}
\acro{EPSRC}{Engineering and Physical Sciences Research Council}
\acro{EQ}{Equalisation}
\acro{ERB}{Equivalent Rectangular Bandwidth}
\acro{ERP}{Ear Reference Point (cf. ITU-T Rec. P.64 1999)}
\acro{ESA}{Early Stage Assessment}
\acro{ESM}{Equivalent Source Method}
\acro{ESPRIT}{Estimation of Signal Parameters via Rotational Invariance Techniques}
\acro{ETAN}{Equivalent Tolerance to Additional Noise} 
\acro{ETSI}{European Telecommunications Standards Institute}
\acro{EURASIP}{European Association for Signal Processing}
\acro{EUSIPCO}{European Signal Processing Conference}
\acro{Eval}{Evaluation\acroextra{ dataset of the \acs{ACE} Challenge}}
\acro{F1}{F1 Score}
\acro{FAR}{False Alarm Rate}
\acro{FastSLAM}[FastSLAM]{FActored Solution To Simultaneous Localization and Mapping}
\acro{FastSLAM2}[FastSLAM]{FActored Solution To \acs{SLAM}} 
\acro{FAU}{Friedrich-Alexander-Universit{\"a}t}
\acro{FB}{Forward-Backward}
\acro{FB2}[FB]{Fullband}
\acro{FBF}{Fixed Beamformer}
\acro{FBSS}{Forward-Backward Spatial Smoothing}
\acro{FCC}{Federal Communications Commission}
\acro{FC-NN}{Fully Connected Neural Network}
\acro{FDM}{Frequency Division Multiplexing}
\acro{FDR}{False Discovery Rate}
\acro{FDR2}[FDR]{Free-Decay Region}
\acro{FEC}{Forward Error Correction}
\acro{FEM}{Finite Element Method}
\acro{FFI}{Norwegian Defence Research Establishment}
\acro{FFT}{Fast Fourier Transform}
\acroindefinite{FFT}{an}{a}
\acro{FIFO}{First-In First-Out}
\acro{FIR}{Finite Impulse Response\acroextra{. A filter whose output is a weighted sum of past input values and whose system function contains only zeros and no poles.}}
\acro{FISM}{Fast Image Source Method}
\acro{FISST}{Finite Set STatistics}
\acro{FLOM}{Fractional Lower-Order Moments}
\acro{FLOS}{Fractional Lower-Order Statistics}
\acro{FM}{Frequency Modulation}
\acro{FMM}{Fast Multipole Method}
\acro{FN}{False Negative}
\acro{FN2}{Nth Formant}
\acro{FNR}{False Negative Rate}
\acro{FORTRAN}{The IBM Mathematical Formula Translating System}
\acro{FoV}{Field of View}
\acro{FP}{False Positive}
\acro{FPR}{False Positive Rate}
\acro{FPS}{Frames Per Second}
\acro{FRI}{Finite Rate of Innovation}
\acro{FSB}{Filter-and-Sum Beamformer}
\acro{FSK}{Frequency Shift Keying}
\acro{FT}{Flat-Top}
\acro{FWER}{Familywise Error Rate}
\acro{FWSSNR}{Frequency-Weighted Segmental \acs{SNR}}
\acro{FWSSRR}{Frequency-Weighted \acs{SSRR}}
\acro{G.711}{\acs{PCM} of Voice Frequencies}
\acro{GARCH}{Generalized Auto-regressive Conditional Heteroscedasticity}
\acro{GBW}{Gain Bandwidth Product}
\acro{GCC}{Generalized Cross-Correlation}
\acro{GCC-PHAT}{Generalized Cross-Correlation with Phase Transform\acroextra{ method of estimating \acs{TDoA}}}
\acro{GCF}{Global Coherence Field}
\acro{GCN}{Graph Convolutional Network}
\acro{GGD}{Generalized Gaussian Distribution}
\acro{GGD2}[G$\Gamma$D]{Generalised Gamma Distribution}
\acro{GM}{Gaussian Mixture}
\acro{GM-PHD}{Gaussian Mixture \acs{PHD}}
\acro{GMCA}{Generalized Morphological Component Analysis}
\acro{GMM}{Gaussian Mixture Model\acroextra{. An approximation to an arbitrary probability density function that consists of a weighted sum of Gaussian distributions}}
\acro{GNN}{Graph Neural Network}
\acro{GNSS}{Global Navigation Satellite System}
\acro{GPRS}{General Packet Radio Services}
\acro{GPS}{Global Positioning System}
\acro{GSC}{Generalized Sidelobe Canceller}
\acro{GSM}{Global System for Mobile Communications}
\acro{GSM-EFR}{\acs{GSM} Enhanced Full Rate Codec}
\acro{GSM-FR}{\acs{GSM} Full Rate Codec}
\acro{GSM-HR}{\acs{GSM} Half Rate Codec}
\acro{GUI}{Graphical User Interface}
\acro{HAAC}{High Amplitude Audio Capture}
\acro{HATS}{Head and Torso Simulator}
\acro{HERB}{Harmonicity-based dEReverBeration}
\acro{HFT}{Hands-Free Terminal}
\acro{HI}{Hearing-Impaired}
\acro{HIE}{Helmholtz Integral Equation}
\acro{HISAS}{High resolution Interferometric \ac{SAS}}
\acro{HINT}{Hearing-in-Noise Test}
\acro{HLT}{Human Language Technology}
\acro{HMM}{Hidden Markov Model}
\acro{HOS}{Higher-Order Statistics}
\acro{HPF}{High-Pass Filter}
\acro{HR}{Half Rate (\acs{GSM} Codec)}
\acro{HRI}{Human-Robot Interaction}
\acro{HRTF}{Head-related Transfer Function}
\acro{HSA}{Hearing, Speech, Audio\acroextra{ technology group at Fraunhofer IDMT}}
\acro{HSD}{Hybrid Steepest Descent}
\acro{HT}{Hannan-Thomson}
\acro{HTK}{Hidden Markov Model Tool Kit}
\acro{IBM}{Ideal Binary Mask}
\acro{IC}{Interference Canceller}
\acro{ICA}{Independent Component Analysis}
\acro{ICASSP}{Intl. Conf. on Acoustics, Speech and Signal Processing}
\acro{ID}{Identifier}
\acro{IDMT}{Institute for Digital Media Technology}
\acro{iDEN}{Integrated Digital Enhanced Network}
\acro{IEC}{International Electrotechnical Commission}
\acro{IEEE}{Institute of Electrical and Electronics Engineers}
\acro{IET}{Institute of Engineering and Technology}
\acro{IETF}{Internet Engineering Task Force}
\acro{IFFT}{Inverse Fast Fourier Transform}
\acro{IHC}{Inner Hair Cell}
\acro{IHT}{Iterative Hard Thresholding}
\acro{IID}{Independent and Identically Distributed}
\acro{IIR}{Infinite Impulse Response\acroextra{. A filter whose output is a weighted sum of both past input and past output values and whose system function contains both poles and zeros.}}
\acro{IL}{Iterated Logarithm\acroextra{, the logarithm of the logarithm}}
\acro{ILAH}{Iterated Logarithm Amplitude Histogram\acroextra{ clipping detection method}}
\acro{ILD}{Interaural Level Difference}
\acro{IMCRA}{Improved Minima Controlled Recursive Averaging\acroextra{. A technique for blindly estimating the spectrum of additive noise in a signal.}}
\acro{IMD}{Inter-Modulation Distortion}
\acro{IMSI}{International Mobile Subscriber Identity}
\acro{IMU}{Inertial Measurement Unit}
\acro{INMD}{In-service Non-intrusive Measurement Device}
\acro{INTERSPEECH}{Annual Conference of the \acs{ISCA}}
\acro{IO}{Infinitely Often}
\acro{IP}{Internet Protocol}
\acro{IPA}{International Phonetic Association}
\acro{IPD}{Interaural Phase Difference}
\acro{IRM}{Ideal Ratio Mask}
\acro{IRS}{Inverse repeated Sequence\acroextra{. A pseudo random sequence used for impulse response measurement.}}
\acro{IRS2}[IRS]{Intermediate Reference System}
\acro{IS}{Importance Sampling}
\acro{ISCA}{International Speech Communication Association}
\acro{ISDN}{Integrated Services Digital Network}
\acro{ISFT}{Inverse \acl{SFT}}
\acro{ISO}{Intl. Organization for Standardization}
\acro{ISP}{Intensity Spectral Profile}
\acro{IST}{Iterative Soft Thresholding}
\acro{ISTFT}{Inverse Short Time Fourier Transform}
\acro{ISVR}{Institute of Sound and Vibration Research\acroextra{, Southampton University, UK}}
\acro{ITD}{Interaural Time Difference}
\acro{ITF}{Interaural Transfer Function}
\acro{ITU}{International Telecommunication Union}
\acro{ITUR}[ITU-R]{International Telecommunication Union\acroextra{ Radiocommunication Sector}}
\acro{ITUT}[ITU-T]{International Telecommunication Union\acroextra{ Telecommunication Standardisation Sector}}
\acro{IUWT}{Isotropic Undecimated Wavelet (Starlet) Transform}
\acro{IWAENC}{Intl. Workshop on Acoustic Signal Enhancement}
\acro{IWAENC_PRE_2012}[IWAENC]{Intl. Workshop Acoustic Echo and Noise Control}
\acro{IWASE}{Intl. Workshop on Acoustic Signal Enhancement}
\acro{JADE}{Joint Approximate Diagonalization of Eigen-Matrices}
\acro{JPDA}{Joint Probabilistic Data Association}
\acro{JPEG}{Joint Photographic Experts Group}
\acro{KDE}{Kernel Density Estimate}
\acro{KF}{Kalman Filter}
\acro{KFD}{Kernel Fisher Discriminant\acroextra{ Analysis}}
\acro{KL}{Karhunen-Lo{\'{e}}ve}
\acro{KLT}{Karhunen-Lo{\'{e}}ve Transform}
\acro{KST}{Kolmogorov-Smirnov Test}
\acro{LAD}{Least Absolute Deviation}
\acro{LAN}{Local Area Network}
\acro{LARS}{Least Angle Regression}
\acro{LAT}[L$_{\textrm{AT}}$]{Equivalent Continuous Sound Level\acroextra{. Also called Leq}}
\acro{LBR}{Low Bitrate Redundancy}
\acro{LC}{Local Criterion}
\acro{LCMP}{Linearly Constrained Minimum Power}
\acro{LCMV}{Linearly Constrained Minimum Variance}
\acro{LCWM}{Linear Combination of Weighted Medians}
\acro{LDC}{Linguistic Data Consortium}
\acro{LEM}{Loudspeaker-Enclosure-Microphone System}
\acro{Leq}[L$_{\textrm{eq}}$]{Equivalent Continuous Sound Level\acroextra{. Also called LAT}}
\acro{LF}{Liljencrants-Fant\acroextra{. The developers of a glottal waveform model}}
\acro{LHS}{Left-Hand Side}
\acro{LID}{Language Identification}
\acro{LILAH}{\acs{LSR2}-\acs{ILAH}\acroextra{ clipping detection method}}
\acro{LIME}{LInear Predictive Multi-input Equalization\acroextra{ algorithm}}
\acro{LiNoPS}{Lightweight Noise Protection System}
\acro{LLN}{Law of Large Numbers}
\acro{LLR}{Log-Likelihood Ratio}
\acro{LLS}{Logarithmic Least Squares}
\acro{LMA}{Least Mean Absolute}
\acro{LMS}{Least Mean Squares\acroextra{ adaptive filter}}
\acro{LNA}{Low Noise Amplifier}
\acro{LoS}{Line-of-Sight}
\acro{LOT}{Listening-Only Test}
\acro{LP}{Linear Parameter}
\acro{LP2}[LP]{Linear Predictive}
\acro{LP3}[LP]{Linear Prediction}
\acro{LPC}{Linear Predictive Coding\acroextra{. An autoregressive model of speech production.}}
\acro{LQO}{Listening Quality Objective}
\acro{LREC}{Conf. on Language Resources and Evaluation}
\acro{LS}{Least-Squares}
\acro{LSA}{Log Spectral Amplitude}
\acro{LSB}{Lower Side-Band}
\acro{LSB2}[LSB]{Least Significant Bit}
\acro{LSD}{Log Spectral Distortion}
\acro{LSP}{Line Spectrum Pairs}
\acro{LSR}{Late Stage Review}
\acro{LSR2}[LSR]{Least Squares Residuals\acroextra{ clipping detection method}}
\acro{LSRT}{Least Squares Residuals with Thresholding\acroextra{ clipping detection method}}
\acro{LSTM}{Long Short-Term Memory}
\acro{LTASS}{Long Term Average Speech Spectrum}
\acro{LTI}{Linear Time Invariant}
\acro{LTP}{Long Term Prediction}
\acro{LU}{Loudness Unit}
\acro{LUFS}{Loudness Units Full-Scale}
\acro{MA}{Moving Average}
\acro{MAD}{Median Absolute Deviation}
\acro{MAE}{Mean Absolute Error}
\acro{MAP}{Maximum \emph{a posteriori}}
\acro{MARDY}{Multichannel Acoustic Reverberation Database at York}
\acro{MARS}{Multivariate Adaptive Regression Splines}
\acro{MBF}{Matched Filter Beamformer}
\acro{MC}{Monte Carlo}
\acro{MCA}{Morphological Component Analysis}
\acro{MCC}{Matthew's Correlation Coefficient}
\acro{MCEQ}{MultiChannel EQualisation}
\acro{MCS}{Multidimensional Colouration Space}
\acro{MDCT}{Modified Discrete Cosine Transform}
\acro{MDL}{Minimum Description Length}
\acro{MDS}{Multidimensional Scaling}
\acro{Mel}{\acroextra{A non-uniform frequency scale corresponding to perceived frequency. It is approximately linear at low frequencies and logarithmic at high frequencies.}}
\acro{MELP}{Mixed Excitation Linear Prediction}
\acro{MFCC}{Mel-frequency Cepstral Coefficients}
\acro{MFS}{Method of Fundamental Solutions}
\acro{MFSK}{Multi-Frequency Shift Keying}
\acro{MHT}{Multi-Hypotheses Tracking}
\acro{MI}{Mutual Information}
\acro{MIMO}{Multiple-Input-Multiple-Output}
\acro{MINT}{Multiple-input/output INverse Theorem}
\acro{MIRS}{Motorola Integrated Radio System}
\acro{MIT}{Massachusetts Institute of Technology}
\acro{MIT-LCS}{Massacchusetts Institute of Technology Laboratory for Computer Science}
\acro{ML}{Maximum Likelihood}
\acro{MLD}{Masking Level Difference}
\acro{MLE}{Maximum Likelihood Estimation}
\acro{ML-TDoA}{Maximum Likelihood Time Difference of Arrival}
\acro{MLMF}{Machine Learning with Multiple Features}
\acro{MLP}{Multi-layer Perceptron}
\acro{MLS}{Maximum Length Sequence\acroextra{ of pseudo random bits.}}
\acro{MMSE}{Minimum Mean Squared Error}
\acro{MMT}{Multiscale Median Transform}
\acro{MNRU}{Modulated Noise Reference Unit}
\acro{MOM}{Mean of Maximum}
\acro{MOS}{Mean Opinion Score}
\acro{MOS-LQO}{Mean Opinion Score - Listening Quality Objective}
\acro{MP}{Matching Pursuit}
\acro{MP3}{\acs{MPEG}-2 Audio Layer III}
\acro{MPEG}{Moving Picture Experts Group}
\acro{MRF}{Markov Random Field}
\acro{MRP}{Mouth Reference Point (cf. ITU-T Rec. P.64 1999)}
\acro{MRT}{Modified Rhyme Test}
\acro{MS}{Minimum Statistics}
\acro{MSB}{Most Significant Bit}
\acro{MSC}{Mean Square Coherence}
\acro{MSE}{Mean Square Error}
\acro{MSN}{Multiple Subscriber Number}
\acro{MSNR}{Maximum \acs{SNR}}
\acro{MTF}{Modulation Transfer Function}
\acro{MTM}{Modified Trimmed Mean}
\acro{MUSCLE}{MeasUred Single-CLustEr}
\acro{MUSHRA}{Multi-stimuli Test with Hidden Reference and Anchor}
\acro{MUSHRAR}{Multi-stimuli Test with Hidden Reference and Anchor for Reverberation}
\acro{MUSIC}{Multiple Signal Classification}
\acro{MVDR}{Minimum Variance Distortionless Response\acroextra{ beamformer}}
\acro{MPDR}{Minimum Power Distortionless Response\acroextra{ beamformer}}
\acro{MWF}{Multi-channel Wiener Filter}
\acro{NAH}{Nearfield Acoustic Holography}
\acro{NB}{Narrowband}
\acro{NCM}{Normalized Coherence Metric}
\acro{NH}{Normal-Hearing}
\acro{NIRA}{Non-Intrusive Room Acoustics}
\acro{NISE}{Non-Intrusive \acs{SNR} estimation}
\acro{NISI}{Non-Intrusive Speech Intelligibility Estimation}
\acro{NISQ}{Non-Intrusive Speech Quality Estimation}
\acro{NIST}{National Institute of Standards and Technology}
\acro{NL}{Noise Level}
\acro{NLA}{Non-Linear Approximation}
\acro{NLMS}{Normalized Least Mean Squares\acroextra{ adaptive filter}}
\acro{NMCFLMS}{Normalized Multichannel Frequency Domain Least Mean Square}
\acro{NMF}{Non-negative Matrix Factorization}
\acro{NOISEX-92}{Database to Study the Effect of Additive Noise on Speech Recognition Systems}
\acro{NOIZEUS}{Noisy Speech Corpus for Evaluation of Speech Enhancement Algorithms}
\acro{NOSRMR}{Normalized Overall \acs{SRMR}}
\acro{NOS}{Number of Sources}
\acro{NOSRMR}{Normalized Overall \acs{SRMR}}
\acro{NPM}{Normalized Projection Misalignment}
\acro{NPV}{Negative Predictive Value}
\acro{NS}{Noise Suppression}
\acro{NSRMR}{Normalised \acs{SRMR}}
\acro{NSRR}{Normalized Signal-to-Reverberation Ratio}
\acro{NSRMR}{Normalised \acs{SRMR}}
\acro{NSV}{Negative-Side Variance}
\acro{NTP}{Network Time Protocol}
\acro{NV}{Noise-Vocoding}
\acro{OBL}{Octave Band Level}
\acro{ODF}{Overdrive Factor}
\acro{Ofcom}{Office of Communications\acroextra{, the independent regulator and competition authority for the UK communications industries}}
\acro{OFDM}{Orthogonal Frequency Division Multiplexing}
\acro{OHC}{Outer Hair Cell}
\acro{OIM}{Objective Intelligibility Measure}
\acro{OLA}{Overlap-add}
\acro{OM-LSA}{Optimally Modified Log-Spectral Amplitude{ Estimator}}
\acro{OMP}{Orthogonal Matching Pursuit}
\acro{OSI}{Open Systems Interconnection}
\acro{OSPA}{Optimal Subpattern Assignment}
\acro{OSRMR}{Overall \acs{SRMR}}
\acro{PAB-SRMR}{Per acoustic band \acs{SRMR}}
\acro{PALM}{Passive Acoustic Localization and Mapping}
\acro{PAMS}{Perceptual Analysis Measurement System}
\acro{PARCOR}{Partial Correlation Coefficients}
\acro{PB}{Phonetically Balanced}
\acro{PBF}{Positive Boolean Function}
\acro{PCA}{Principal Components Analysis}
\acro{PDA}{Personal Digital Assistant}
\acro{PDA2}[PDA]{Probabilistic Data Association}
\acro{PDE}{Partial Differential Equation}
\acro{pdf}{Probability Density Function}
\acro{PDF}{Probability Density Function}
\acro{PE}{Parameter Estimation}
\acro{PEASS}{Perceptual Evaluation for Audio Source Separation}
\acro{PEFAC}{Pitch Estimation Filter with Amplitude Compression}
\acro{PESQ}{Perceptual Evaluation of Speech Quality}
\acro{PF}{Psychometric Function}
\acro{pgfl}[p.g.fl.]{Probability Generating Functional}
\acro{PHAT}{Phase Transform}
\acro{PHD}{Probability Hypothesis Density}
\acro{PIP}{Peak-Image Pairing}
\acro{PIV}{Pseudo-Intensity Vector}
\acro{PL}{Pseudo-likelihood}
\acro{PLC}{Packet Loss Concealment}
\acro{PLL}{Phase Locked Loop}
\acro{PLP}{Perceptual Linear Prediction}
\acro{PLR}{Perceived Level of Reverberation}
\acro{PM}{Phase Modulation}
\acro{PMF}{Probability Mass Function}
\acro{PMOS}{Predicted Mean Opinion Score}
\acro{PNN}{Parameterized Neural Network}
\acro{POLQA}{Perceptual Objective Listening Quality Analysis}
\acro{POTS}{Plain Old Telephone Service}
\acro{PPP}{Poisson Point Process}
\acro{PPS}{Pulse-Per-Second}
\acro{PPV}{Positive Predictive Value}
\acro{PReLU}{Parametric Rectified Linear Unit}
\acro{PRLM}{Phoneme Recognition and Language Modelling}
\acro{PRP}{Pair-wise Relative Phase-ratio}
\acro{PSD}{Power Spectral Density}
\acro{PSK}{Phase Shift Keying}
\acro{PSNR}{Peak Signal-to-Noise Ratio}
\acro{PSOLA}{Pitch Synchronous Overlap Add\acroextra{. A method of scaling a signal in time and pitch independently.}}
\acro{PSQM}{Perceptual Speech Quality Measurement}
\acro{PSSL}{Positional Sound Source Localization}
\acro{PSTN}{Public Switched Telephone Network}
\acro{PWD}{Plane-Wave Decomposition}
\acro{PTA}{Pure-Tone Audiology}
\acro{QAM}{Quadrature Amplitude Modulation}
\acro{QILAH}{Quadrisected Iterated Logarithm Amplitude Histogram\acroextra{ clipping detection method}}
\acro{QMF}{Quadrature Mirror Filter}
\acro{QoE}{Quality-of-Experience}
\acro{QoS}{Quality-of-Service}
\acro{QPSK}{Quadrature Phase Shift Keying}
\acro{RADAR}{RAdio Detection And Ranging}
\acro{RASTA}{Relative Spectral}
\acro{RASTA-PLP}{Relative Spectral Perceptual Linear Prediction}
\acro{RASTI}{Room Acoustics Speech Transmission Index\acroextra{ (superseded by STIPA)}}
\acro{RBM}{Restricted Boltzmann Machine}
\acro{RB-PHD}{Rao-Blackwellised \acs{PHD}}
\acro{RBPF}{Rao-Blackwellised Particle Filter}
\acro{RC}{Relative Criterion}
\acro{RDT}[$R_\textrm{DT}$]{Reverberation Decay Tail}
\acro{RDTF}{Relative Direct Transfer Function}
\acro{ReLU}{Rectified Linear Unit}
\acro{RF}{Radio Frequency}
\acro{RFI}{Radio Frequency Interference}
\acro{RFS}{Random Finite Set}
\acro{RHS}{Right-Hand Side}
\acro{RIP}{Restricted Isometry Property}
\acro{RIR}{Room Impulse Response}
\acro{RLS}{Recursive Least Squares\acroextra{ adaptive filter}}
\acro{RLSD}{Relative Log Spectral Distortion}
\acro{RMCLS}{Relaxed Multichannel Least Squares}
\acro{RMCLS-CIT}{Relaxed MultiChannel Least-Squares with Constrained Initial Taps}
\acro{RMS}{Root Mean Square}
\acro{RMSE}{Root Mean Square Error}
\acro{RNN}{Recurrent Neural Network}
\acro{RelNet}{Relation network}
\acro{ROC}{Receiver Operating Characteristic}
\acro{ROHC}{Robust Header Compression}
\acro{RP}{Received Pronunciation}
\acro{RPE}{Regular Pulse Excitation}
\acro{RRTF}{Relative Real-Time Factor}
\acro{RS}{Reverberation Suppression}
\acro{RSM}{Reflector Source Method}
\acro{RSS}{Received Signal Strength}
\acro{RSV}{Room Spectral Variance}
\acro{RT}{Reverberation Time}
\acro{RTAN}{Robustness to Additional Noise}
\acro{RTF}{Real-Time Factor}
\acro{RTF2}[RTF]{Room Transfer Function}
\acro{RTF3}[RTF]{Relative Transfer Function}
\acro{RV}{Random Variable}
\acro{RVP}{Recursive Vector Projection}
\acro{RWTH}{Rheinisch-Westf\"{a}lische Technische Hochschule}
\acro{S50}[$S_{50}$]{Intelligibility Function Gradient at the \ac{SRT}}
\acro{SA}{Spectral Amplitude}
\acro{SAA}{Synthetic Aperture Audio}
\acro{SAP}{Speech And Audio Processing}
\acro{SAR}{Speech-to-Artifact Ratio}
\acro{SAR2}[SAR]{Speaker Alternation Rate}
\acro{SAR3}[SAR]{Synthetic Aperture \ac{RADAR}}
\acro{SAS}{Synthetic Aperture \ac{SONAR}}
\acro{SB}{Subband}
\acro{SC-PHD}{Single Cluster \acs{PHD}}
\acro{SCAF}{Single Channel Adaptive Filter}
\acro{SCB}{Stochastic Codebook}
\acro{SCNR}{Single-Channel Noise Reduction}
\acro{SCOT}{Smoothed Coherence Transform}
\acro{SCNR}{Single-Channel Noise Reduction}
\acro{SC-PHD}{Single Cluster \acs{PHD}}
\acro{SCR}{Signal-to-Competition Ratio}
\acro{SCRIBE}{Spoken Corpus of British English}
\acro{SCT}{Speech Corruption Toolkit}
\acro{SCT2}{Short Conversation Test}
\acro{SD}{Semantic Differential}
\acro{SDB}{Superdirective Beamformer}
\acro{SDD}{Spectral Decay Distributions}
\acro{SDDMSB}{\acs{SDD} with Mel-spaced frequency bands}
\acro{SDDSA}{\acs{SDD} with Mel-spaced frequency bands and selective averaging}
\acro{SDDSA-G}{\acs{SDDSA} with Gerkmann noise estimator}
\acro{SDDSA-H}{\acs{SDDSA} with Hendriks noise estimator}
\acro{SDR}{Software Defined Radio}
\acro{SDR2}{Speech}
\acro{SDRAM}{Synchronous Dynamic Random Access Memory}
\acro{SDT}{Speech Description Taxonomy}
\acro{SEDF}{Subband \ac{EDF}}
\acro{SELD}{Sound Event Localization and Detection}
\acro{SEMG}{Surface Electromyography}
\acro{SFDR}{Spurious Free Dynamic Range}
\acro{SFM}{Single Feature with Mapping}
\acro{SFT}{Spherical Fourier Transform}
\acro{SH}{Spherical Harmonic}
\acro{SHD}{Spectral Harmonic Decomposition}
\acro{SHD2}[SHD]{Spherical Harmonic Domain}
\acro{SIE}{System Identification Error}
\acro{SII}{Speech Intelligibility Index}
\acro{SIImod}{Speech Intelligibility Index in the modulation domain}
\acro{SIM}{Subscriber Identity Module}
\acro{SIMO}{Single-Input-Multiple-Output}
\acro{SINAD}{Signal-to-Noise and Distortion Ratio}
\acro{SIP}{Session Initiation Protocol}
\acro{SIR}{Signal-to-Interference Ratio}
\acro{SIR2}[SIR]{Sequential Importance Resampling}
\acro{SIREAC}{Simulation of REal Acoustics\acroextra{ Software Tool}}
\acro{SIS}{Sequential Importance Sampling}
\acro{SL}{Speech Level}
\acro{SLAM}{Simultaneous Localization and Mapping}
\acro{SLLN}{Strong Law of Large Numbers}
\acro{SLM}{Sound Level Meter}
\acro{SLF}{Spatial Likelihood Function}
\acro{SMA}{Spherical Microphone Array}
\acro{SMARD}{Single- and Multichannel Audio Recordings Database}
\acro{SMERSH}{Spatiotemporal Averaging Method for Enhancement of Reverberant Speech}
\acro{SMIR}{Spherical Microphone array Impulse Response}
\acro{SMPTE}{Society of Motion Picture and Television Engineers}
\acro{SMS}{Short Message Service}
\acro{SNR}{Signal-to-Noise Ratio}
\acro{SNR2}[SNR]{Speech-to-Noise Ratio}
\acro{SNT}{Subspace Noise Tracking\acroextra{ algorithm}}
\acro{SOBM}{STOI-optimal Binary Mask}
\acro{SONAR}{SOund Navigation And Ranging}
\acro{SOLA}{Synchronous Overlap Add\acroextra{. A method of scaling a signal in time and pitch independently.}}
\acro{SPC}{Specificity}
\acro{SPEECON}{Speech Databases for Consumer Devices}
\acro{SPHERE}{NIST SPeech Header REsources\acroextra{ software with embedded Shorten Compression}}
\acro{SPIN}{Speech Perception In Noise}
\acro{SPL}{Sound Pressure Level}
\acro{SPP}{Speech Presence Probability}
\acro{SPQA}{Speech Quality Assurance Package}
\acro{SQNR}{Signal-to-Quantization Noise Ratio}
\acro{SR}{Sparse Representation}
\acro{SR2}[SR]{Spectral Rotation}
\acro{SRA}{Statistical Room Acoustics}
\acro{SRI}{SRI International\acroextra{. Formerly Standford Research Institute}}
\acro{SRMR}{Speech-to-Reverberation Modulation Energy Ratio}
\acro{SRP}{Steered Response Power}
\acro{SRP-PHAT}{Steered Response Power with Phase Transform}
\acro{SRP-TDE}{Steered Response Power with Time Delay Estimation}
\acro{SRR}{Signal-to-Reverberation Ratio}
\acro{SRT}{Speech Reception Threshold\acroextra{ (also known as Speech Recognition Threshold)}}
\acro{SS}{Spectral Subtraction}
\acro{SSB}{Single Side-Band}
\acro{SSI}{Synthetic Sentence Identification}
\acro{SSL}{Sound Source Localization}
\acro{SSN2}[SSN]{Simultaneous Switching Noise}
\acro{SSN}{Speech-Shaped Noise}
\acro{SSNR}{Segmental \acs{SNR}}
\acro{SSOBM}{Stochastic \acs{SOBM}}
\acro{SSRR}{Segmental Signal-to-Reverberation Ratio}
\acro{SSW}{Staggered Spondaic Word}
\acro{STFT}{Short Time Fourier Transform}
\acro{STI}{Speech Transmission Index}
\acro{STIPA}{Speech Transmission Index for Public Address Systems}
\acro{STITEL}{Speech Transmission Index for Telecommunication Systems}
\acro{STMI}{Spectro-Temporal Modulation Index}
\acro{STNR}{\acs{NIST}'s Speech-to-Noise Ratio\acroextra{ Estimation Algorithm}}
\acro{STOI}{Short-Time Objective Intelligibility\acroextra{ measure}}
\acro{STQ}{Speech Processing, Transmission and Quality Aspects}
\acro{STSA}{Short Time Spectral Analysis}
\acro{STSA1}[STSA]{Short Time Spectral Amplitude}
\acro{SUS}{Semantically Unpredictable Sentences}
\acro{SVD}{Singular Value Decomposition}
\acro{SVM}{Support Vector Machine}
\acro{SWSOBM}{Stochastic \acs{WSOBM}}
\acro{T20}[$T_\textrm{20}$]{Reverberation Time\acroextra{ to decay by $20$ dB}}
\acro{T30}[$T_\textrm{30}$]{Reverberation Time\acroextra{ to decay by $30$ dB}}
\acro{T60}[$T_\textrm{60}$]{Reverberation Time\acroextra{ to decay by $60$ dB}}
\acro{TBM}{Target Binary Mask}
\acro{TDE}{Time Delay Estimation}
\acro{TDHS}{Time Domain Harmonic Scaling\acroextra{. A method of scaling a signal in time and pitch independently.}}
\acrodefplural{TDOA}[TDOAs]{Time-Differences-of-Arrival}
\acro{TDOA}{Time-Difference-of-Arrival}
\acrodefplural{TDOA}[TDOAs]{Time-Differences-of-Arrival}
\acro{TDT}{Tone Decay Test}
\acro{TF}{Time-Frequency}
\acro{TFS}{Temporal Fine Structure}
\acro{TFGM}{Time-Frequency Gain Modification\acroextra{. An approach to signal enhancement in which a signal is multiplied by a gain function in the time-frequency domain.}}
\acro{THD}{Total Harmonic Distortion}
\acro{TI}{Texas Instruments, Inc.}
\acro{TIMIT}{\acs{TI}-\acs{MIT} speech corpus}
\acro{TIPHON}{Telecommunication and Internet Protocol Harmonization Over Networks}
\acro{TLS}{Total Least-Squares}
\acro{TN}{True Negative}
\acro{TNR}{True Negative Rate}
\acro{TOA}{Time-of-Arrival}
\acrodefplural{TOA}[TOAs]{Times-of-Arrival}
\acro{TOF}{Time-of-Flight}
\acro{TOSQA}{Telekom Objective Speech Quality Assessmentt}
\acro{TP}{True Positive}
\acro{TP2}[TP]{Trivial Pursuit}
\acro{TPCC}{Trivial Pursuit with Clipping Constraints}
\acro{TPR}{True Positive Rate}
\acro{TSE}{Taylor Series Expansion}
\acro{TVAR}{Time-varying Autoregression}
\acro{UAV}{Unmanned Aerial Vehicle}
\acro{UDP}{User Datagram Protocol}
\acro{UFRJ}{Federal University of Rio de Janeiro}
\acro{UHF}{Ultra High Frequency}
\acro{UKF}{Unscented Kalman Filter}
\acro{ULA}{Uniform Linear Array}
\acro{ULF}{Ultra Low Frequency}
\acro{UMTS}{Universal Mobile Telecommunications Service}
\acro{US}{United States}
\acro{UTBM}{Universal Target Binary Mask}
\acro{VAD}{Voice Activity Detector}
\acro{VBR}{Variable Bit-Rate}
\acro{VCV}{Vowel-Consonant-Vowel}
\acro{VRD}{Variance of Decay-rates}
\acro{VGC}{Voice Grade Channel}
\acro{vMF}{von Mises-Fisher}
\acro{VoIP}{Voice Over Internet Protocol}
\acro{VRT}{Vlaamse Radio- en Televisieomroeporganisatie\acroextra{. (Flemish Radio and Television Broadcasting Organization)}}
\acro{VSELP}{Vector Sum-excited Linear Prediction}
\acro{VST}{Virtual Studio Technology\acroextra{. An interface standard developed by Steinberg for adding plugins to an audio editor.}}
\acro{WADA}{Waveform Amplitude Distribution Analysis}
\acro{WASN}{Wireless Acoustic Sensor Network}
\acrodefplural{WASN}[WASNs]{Wireless Acoustic Sensor Networks}
\acro{WASPAA}{Workshop on Applications of Signal Processing to Audio and Acoustics}
\acro{WAV}{Waveform Audio File Format}
\acro{WAVE}{Waveform Audio File Format}
\acro{WB}{Wideband}
\acro{WER}{Word Error Rate}
\acro{WGN}{White Gaussian Noise}
\acro{WLAN}{Wireless \acs{LAN}}
\acro{WLLN}{Weak Law of Large Numbers}
\acro{WM}{Working Memory}
\acro{WMA}{Windows Media Audio}
\acro{WNG}{White Noise Gain}
\acro{WSS}{Weighted Spectral Slope}
\acro{WSOBM}{Weighted \acs{SOBM}}
\acro{WSTOI}{Weighted \acs{STOI}}
\acro{ZOS}{Zero-Order Statistics}
\end{acronym}

\title{Controlling the Parameterized Multi-channel Wiener Filter using a tiny neural network}

\name{\begin{minipage}{\textwidth}\centering
Eric Grinstein$^1$\thanks{Work done during research internship at Meta},
Ashutosh Pandey$^2$,
Cole Li$^2$,
Shanmukha Srinivas$^3$ \\
Juan Azcarreta$^2$,
Jacob Donley$^2$,
Sanha Lee$^2$,
Ali Aroudi$^2$,
Çağdaş Bilen$^2$
\end{minipage}}
\address{$^{1}$Imperial College London, U.K. \\
$^{2}$Meta Reality Labs \\
$^{3}$Ohio State University
}

\begin{document}
\maketitle

\begin{abstract}
    Noise suppression and speech distortion are two important aspects to be balanced when designing multi-channel \ac{SE} algorithms. Although neural network models have achieved state-of-the-art noise suppression, their non-linear operations often introduce high speech distortion. Conversely, classical signal processing algorithms such as the \ac{PMWF} beamformer offer explicit mechanisms for controlling the suppression/distortion trade-off. In this work, we present NeuralPMWF, a system where the \ac{PMWF} is entirely controlled using a low-latency, low-compute neural network, resulting in a low-complexity system offering high noise reduction and low speech distortion. Experimental results show that our proposed approach results in significantly better perceptual and objective speech enhancement in comparison to several competitive baselines using similar computational resources.
\end{abstract}

\vspace{0.3cm}
\section{Introduction}

\ac{MCSE} and beamforming are fundamental tasks in audio signal processing. They consists of estimating a target speaker's sound from a set of noisy microphone array measurements. Example applications of \ac{MCSE} are voice capture systems, such as voice assistants, tele-conference applications, hearing aids and wearables, such as smart glasses \cite{Haykin2009, Guiraud2022}. Effective \ac{SE} methods must remove noise while preserving speech inteligibility and quality, that is, without introducing speech distortion \cite{Souden2010}. This task is particularly challenging under environments containing low \ac{SNR}, moving sources, babble noise, reverberation, among other effects.

\ac{SE} methods may be categorized as \ac{DSP}-based, \ac{DNN}-based, or hybrid. Many \ac{DSP}-based methods are typically based on linear filters, and received significant attention for more than five decades \cite{Ephraim1984, Benesty2008, Tashev2009, Gannot2017}. An important aspect when choosing a beamformer is its noise suppression vs speech distortion trade-off. A popular high suppression/distortion beamformer is the \ac{MWF} \cite{VanTrees2002, Wang2023}, while popular low distortion ones are the \ac{MVDR} beamformer \cite{Capon1969, Griffiths1982, Lorenz2005} and the \ac{GSC} \cite{Buckley1986}. This work gives particular focus to the \acf{PMWF} \cite{Souden2010, Doclo2007}, a generalization of the three aforementioned beamformers which allows for the beamformer's suppression/distortion pattern to be controlled for each frequency band. Although dynamically controlling this pattern offers great potential in making the filter adapt to different scenarios, the only works we found to do so were \cite{Braun2015, Bagheri2019, Ngo2009}, which rely on specialized parameter tuning, which hinders the method's applicability in practical scenarios.

Conversely, many \ac{DNN}-based methods have also been proposed for multi-channel \ac{SE} \cite{Wang2018, Tolooshams2020, Liu2020, Lee2023}, where multiple non-linear operations result in excellent noise suppression at the cost of speech distortion. Furthermore, neural methods often incur high memory and CPU usage, making them unsuitable for portable devices such as smart glasses. This paper focuses on hybrid methods \cite{Hsieh2024, Tao2024, Pandey2022, Wang2023, Tammen2023}, which jointly utilize \ac{DSP} and \ac{DNN} modules, allowing for systems with better noise suppression than pure \ac{DSP} and lower speech distortion and complexity than end-to-end \ac{DNN} systems.

\begin{figure}
    \centering
    \includegraphics[width=\columnwidth]{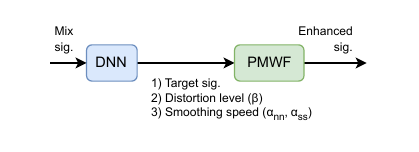}
    \caption{Overview of the proposed NeuralPMWF \ac{MCSE} method.}
    \label{fig:simplified_system_overview}
\end{figure}

As shown in \autoref{fig:simplified_system_overview}, we propose NeuralPMWF, a hybrid \ac{MCSE} model employing a small \ac{DNN} to fully control the \ac{PMWF}. The \ac{PMWF} requires estimation of speech and noise covariance matrices, which have been classically estimated using \ac{DSP} methods, usually involving \ac{SPP} \cite{Souden2010b}. In contrast, we estimate these statistics from a multi-channel target signal estimate produced by a \ac{DNN}, from which covariance matrices are estimated using exponential smoothing. Our system also estimates the required parameters $\alpha_{ss}$ and $\alpha_{nn}$ for smoothing, which control the speed to which the filter adapts to acoustic changes. Finally, our filter dynamically controls the \ac{PMWF}'s distortion parameter $\beta$, which we show to result in a significant performance improvement.

The main contribution of this paper is therefore a novel low-latency and low-compute multi-channel speech enhancement system using a small neural network and the \ac{PMWF}, along with two ablation studies, and comparisons with state-of-the-art baselines. This paper continues as follows. In \autoref{sec:signal_model}, the signal model and system goal are described. In \autoref{sec:method}, the proposed NeuralPWMF method is described. In \autoref{sec:experimentation}, we perform an extensive ablation study of the proposed approach, followed by a comparison with competitive baselines. Finally, we analyze the results in \autoref{sec:results} and conclude the work in \autoref{sec:conclusion}.

\begin{figure*}[h]
    \centering
    \includegraphics[width=\textwidth]{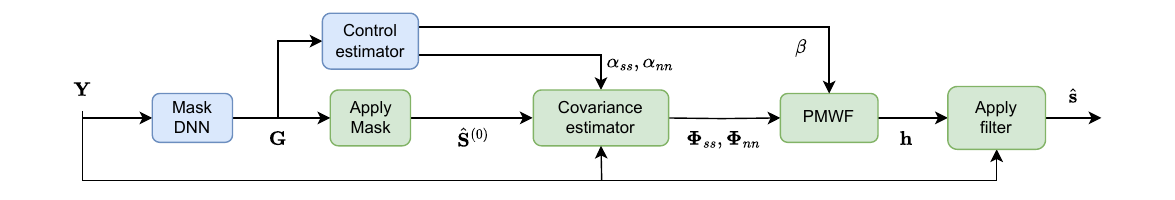}
    \caption{System overview. DNN blocks are coloured in blue, and DSP blocks are coloured in green.}
    \label{fig:system_overview}
\end{figure*}

\vspace{0.3cm}
\section{Signal model, notation and goal} \label{sec:signal_model}
We represent the multi-channel speech signal as $\mathbf{S}$, the received signal by the microphone array as $\mathbf{Y}$, and the additive multi-channel noise, $\mathbf{N}$, all represented as tensors in \ac{STFT} domain as
\begin{equation} \label{eq:sig-model-tensor}
   \mathbf{Y} = \mathbf{S} + \mathbf{N} \in \mathbb{C}^{T \times F \times M} ,
\end{equation}
where $T,F$ are the total number of time and frequency bins of interest, and $M$ is the number of available microphone channels. Note that although all time bins are expressed in \eqref{eq:sig-model-tensor}, the systems analyzed in this paper are causal, processing each time bin using past information only. We use a programming-inspired tensor indexing to access specific time/frequency/channel bins in \eqref{eq:sig-model-tensor}. Namely, we define the goal of a multi-channel \ac{SE} system as estimating the source signal at reference channel $0$, that is, $\mathbf{S}_0 = \mathbf{S}[:, :, 0]$, where $:$ refers to all elements within the axis of interest. We shall also use the shorthand $\mathbf{X}[t, w] = \mathbf{X}[t, w, :]$ when accessing all microphone channels of particular \ac{TF} bins of $\mathbf{X}$.

\vspace{0.3cm}
\section{Proposed method: Neural PMWF} \label{sec:method}

The system diagram is shown in \autoref{fig:system_overview}. The goal of our system is to estimate a \ac{TF} varying filter $\mathbf{h}[t, w] \in \mathbb{C}^{M}$ that defines a combination of all channels signals, that is
\begin{equation} \label{eq:filt}
    \hat{s}[t, w] = \mathbf{h}[t, w]^T \mathbf{S}[t, w].
\end{equation}
The motivation of employing linear filtering in favour of end-to-end neural systems is introducing meaningful signal processing-based inductive biases, which result in lower speech distortion levels on smaller networks. In our application, the filter $\mathbf{h}[t, w]$ \eqref{eq:filt} is the \ac{PMWF}, which was chosen for its explicit ability of controlling speech distortion. It is derived in \cite{Souden2010} using Lagrange multipliers as
\begin{equation} \label{eq:pmwf}
\begin{split}
    \pmb{\gamma}[t, w] &= \mathbf{\Phi}^{-1}_{nn}[t, w] \mathbf{\Phi}_{ss}[t, w] \in \mathbb{C}^{M \times M} \\
    \mathbf{h}[t, w] &= \frac{\pmb{\gamma}[t, w][:, 0]}{\beta[t, w] + \text{trace}(\pmb{\gamma}[t, w])},
\end{split}
\end{equation}
Where variable $\pmb{\gamma}$ is defined to improve the formula's legibility. We first note that the formula requires the speech and noise covariance matrices, defined as $\mathbf{\Phi}_{xx} = \mathbb{E}(\mathbf{X}[t, w]\mathbf{X}[t, w]^H ) \in \mathbb{C}^{M \times M}$, where $\mathbf{X} = \{ \mathbf{S}, \mathbf{N}\}$. Latent signals $\mathbf{S}$ and $\mathbf{N}$ are estimated from mixture using a \ac{DNN} described in \autoref{sec:dnn}. The numerator $\pmb{\gamma}[t, w][:, 0]$  corresponds to choosing reference channel $0$ of $\pmb{\gamma}$. The denominator is a sum of the $\text{trace}$, i.e., the sum of $\pmb{\gamma}$'s diagonal elements with the parameter $\beta$, which controls the filter's output speech distortion and is central to this work.

An essential step for computing the filter coefficients using \eqref{eq:pmwf} consists of estimating the speech and noise covariance matrices. In this work, we use a network to learn a multi-channel complex-valued \ac{TF} mask, which is then used to estimate a multi-channel target speech signal $\hat{\mathbf{S}}_0 \in \mathbb{C}^{T \times F \times M}$, which is then used to estimate noise as $\hat{\mathbf{N}}_0 = \mathbf{Y} - \hat{\mathbf{S}}_0$ (using \eqref{eq:sig-model-tensor}). Finally, $\hat{\mathbf{S}}_0$ and $\hat{\mathbf{N}}_0$ are used to obtain the final covariance matrices through exponential smoothing as
\begin{equation} \label{eq:smoothing}
\begin{split}
    \mathbf{\Phi}_{ss}[t,w] = \; & (1 - \alpha_{ss}[w])\mathbf{\Phi}_{ss}[t - 1, w] \\
                                &+ \alpha_{ss}[w](\hat{\mathbf{S}}_0[t,w]\hat{\mathbf{S}}_0[t,w]^H) \\
    \mathbf{\Phi}_{nn}[t,w] = \; & (1 - \alpha_{nn}[w])\mathbf{\Phi}_{ss}[t - 1, w] \\
                                &+ \alpha_{nn}[w](\hat{\mathbf{N}}_0[t,w]\hat{\mathbf{N}}_0[t,w]^H),
\end{split}
\end{equation}
where $0 < \alpha_{ss},\alpha_{nn} < 1$ control the filter's adaptation speed. In our work, we learn optimal frequency-dependent $\alpha$s during training, as we found it advantageous to allow some bins to change faster than others. Works such as \cite{Cohen2001, Cohen2003, Souden2010b} propose to make the $\alpha$'s \ac{TF} varying, which we include as model variants in our ablation studies. Finally, many works set the distortion control parameter $\beta$ as $0$ or $1$ as to obtain the classical \ac{MVDR} and \ac{MWF} filters respectively. In contrast, we will show that dynamically controlling $\beta$ results in expressive performance gains. The proposed strategies for controlling $\alpha$ and $\beta$ are described below.

\subsection{Control estimation} \label{sec:controls}

\begin{figure*}
    \centering
    \includegraphics[width=\textwidth]{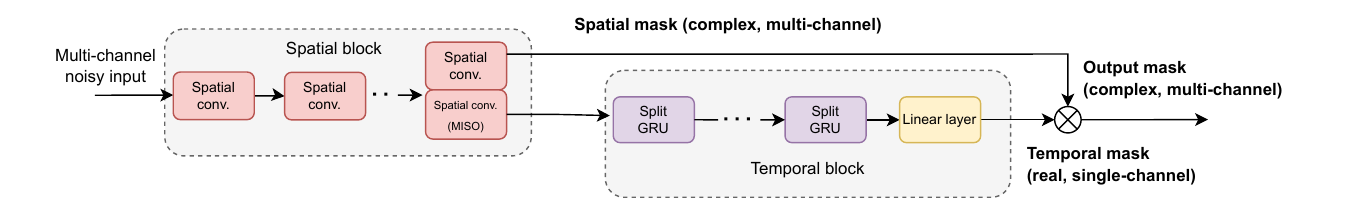}
    \caption{Proposed network architecture for the MaskDNN block in \autoref{fig:system_overview}, which
    estimates the complex-valued mask $\mathbf{G}$.}
    \label{fig:network_architecture}
\end{figure*}

Our proposed method operates in two steps. First, a \ac{DNN} estimates a multi-channel target mask $\mathbf{G}$, which is used both for estimating target signal $\hat{\mathbf{S}}_0$ and filter distortion controls $\mathbf{\beta}$. Furthermore, $\mathbf{\alpha}_{ss}$ and $\mathbf{\alpha}_{nn}$ are jointly learned during training, not requiring mask $\mathbf{G}$ for their estimation. We define the controls $\alpha_{nn}$, $\alpha_{ss}$ and $\beta$ as
\begin{equation} \label{eq:controls}
\begin{split}
     \hat{p}[t, w] &= \text{sigmoid}(\mathbf{p}^{(a)}[w]\lvert \mathbf{G}[t, w, 0] \rvert + \mathbf{p}^{(b)}[w]) \\
     \mathbf{\beta}[t, w] &= \mathbf{\beta}^{(0)}[w](1 - \hat{\mathbf{P}}[t, w])\\
     \mathbf{\alpha}_{ss}[w] &= \text{sigmoid}(\mathbf{\alpha}^{(0)}_{ss}[w]) \\
     \mathbf{\alpha}_{nn}[w] &= \text{sigmoid}(\mathbf{\alpha}^{(0)}_{nn}[w]).
\end{split}
\end{equation}
$\hat{p}$ can be interpreted as the \acf{SPP} \cite{Souden2010b}, which is frequently used as an intermediary feature in \ac{SE}. It is computed using the mask applied to reference channel $0$. $\mathbf{p}^{(a)}, \mathbf{p}^{(b)}, \mathbf{\alpha}^{(0)}_{nn}, \mathbf{\alpha}^{(0)}_{ss}, \mathbf{\beta}^{(0)}_{ss} \in \mathbb{R}^F$ are modeled to act as time-invariant gains to each control parameter, and are learned during training along with the mask network. Note that the operations in \eqref{eq:controls} are not to be confused with fully connected layers, which were found to be more computationally expensive and unstable during training.

\subsection{Network architecture} \label{sec:dnn}

The network is divided in two blocks, the spatial and temporal processing blocks, which have the goals of extracting patterns in channel and time dimensions respectively.

As a first step, the real and imaginary parts of the complex-valued spectrogram are decoupled into separate channels, resulting in a real-valued tensor of dimensions ${2M \times T \times F}$. This tensor is then processed using a sequence of convolutional layers called spatial processing blocks, introduced in \cite{Pandey2024, Pandey2024a}, which mimic a frequency domain Filter-And-Sum \cite{Tashev2009} beamformer operation. Each spatial convolution layer $l$ comprises $F$ distinct matrices, which are multiplied by the input tensor at each frequency bin. Each convolution is followed by parametric ReLU non-linearity. All but the last set of spatial convolutions have the same input/output size except for the last one, which outputs an additional channel which is used as the temporal block inputs. This is expressed in \autoref{fig:network_architecture} as two separate blocks to improve comprehension.

The temporal block consists of a sequence of SplitGRU units introduced in\cite{tan2019learning}, which provide an efficient alternative to classical \ac{GRU} \cite{Chung2014} units. The additional channel generated from the last spatial processing layer is encoded to a feature space of size $H$ by applying a linear layer. These features are directed into a series of causal (uni-directional) SplitGRU layers, each with a hidden size of $H$. The SplitGRU layer differs from conventional GRU layers by dividing the input into $R$ segments across the feature dimension and processing each segment with one of $R$ parallel GRUs. The outputs of each layer are reorganized such that the output from a specific GRU is distributed to all GRUs in the subsequent layer. Employing a split factor of $R$ effectively reduces computational demands by a factor of $R$. The final output of the GRU model is then projected to a size of $F$ using a linear layer.

Finally, the multi- and single-channel outputs from the spatial and temporal blocks are combined into a multi-channel, complex-valued mask $\mathbf{G}$ by treating the temporal block's output as a real-mask itself, which is independently applied to each channel of the spatial block's output.

\vspace{0.3cm}
\section{Experimentation} \label{sec:experimentation}
We perform three experiments, two of which are ablation studies and the remaining one is a comparison of our best performing model and with baselines. The multichannel time-domain signal is normalized to a range between -60~dB and -20~dB at the reference microphone. All models are trained using Pytorch \cite{Paszke2019} with a time-domain \ac{SNR} loss and a frequency-domain phase constrained magnitude (PCM) loss proposed in \cite{Pandey2021}. Training spans $100$ epochs with $10$-~second utterances and a batch size of 128, distributed across multiple Nvidia H100 GPUs. We employ the Adam optimizer \cite{Kingma2014}, with AMSGrad \cite{Reddi2018}, and clipped gradient norms at 1. The learning rate starts at 0.001 for the first 70 epochs and is reduced by a factor of 10\% every 10 epochs thereafter. Model evaluations utilized were \ac{STOI} \cite{Taal2011}, narrow-band \ac{PESQ} \cite{Rix2001}, and SNR w.r.t. the reverberant target speech \cite{Vincent2006}.

The MaskDNN architecture used consists of four spatial convolution filters, where the first 3 maintain the input channel size at 10, representing the real and imaginary parts of the spatial features. The temporal block contains 3 SplitGRU layers with 96 hidden units and 2 splits each. STFT settings include 256-sample window size, 128-sample frame-shift, and 129 frequency bins, therefore characterizing the NeuralPMWF as having an algorithmic latency of $16$~ms to process one audio frame at 16kHz \cite{Wang2023}.

\vspace{0.2cm}
\subsection{Dataset}

To create pairs of clean and noisy signals for training, we utilize the Interspeech 2020 DNS Challenge corpus \cite{Reddy2020}. The speakers and noises in the training set are randomly allocated into training, testing, and validation subsets with split ratios of 85\%, 5\%, and 10\%, respectively. All utterances are resampled to 16 kHz prior to data generation.
Initially, random room dimensions are determined: length and width are uniformly sampled from [3, 10] meters, and height
from [2, 5] meters. We then randomly select the location and orientation of the microphone within the room. The speech source is positioned at a distance ranging from [0.5, 2.5] meters from the array center within a Field-of-View (FoV) of $[-30^{\circ}, 30^{\circ}]$ azimuth and $[-90^{\circ}, 90^{\circ}]$ elevation, with the goal of evaluating performance on smart glasses applications where the source of interest lies in front of the wearer. Between 1 and 10 noise sources are placed at distances exceeding 0.5 meters from the array center. Between 0 and 10 interfering talker locations are
positioned at distances greater than 3 meters. The Signal-to-Noise Ratio (SNR) is sampled from [-5, 10] dB, and the
Signal-to-Interference Ratio (SIR) from [5, 10] dB. Data is simulated using the image method \cite{Allen1979} using Pyroomacoustics \cite{Scheibler2018} with order 6 and a uniformly sampled wall absorption coefficient from [0.1, 0.7]. To simulate multichannel data, we auralize the simulated sources to a 5-microphone array resembling Rayban Meta smart glasses by utilizing measured anechoic \acp{ATF}, simulating the effects of channel directivity, head diffraction and absorption.
In total, we simulate 320k, 600 and 3.2k utterances for training, validation, and testing respectively.

\vspace{0.2cm}
\subsection{Ablation studies: controlling smoothing and distortion}

To obtain the model described in \autoref{sec:method}, we performed two ablation studies focused on estimating the best strategies for controlling covariance smoothing ($\alpha$) and distortion ($\beta$), whose results are shown in \autoref{tab:ablation}. For the $\alpha$ ablation study, we test four different smoothing techniques: (i) cumulative mean \cite{Lorenz2005}, (ii) frequency-independent untrained exponential smoothing, (iii) frequency-dependent trained smoothing and (iv) \ac{SPP}-driven exponential smoothing \cite{Cohen2001} (as described in \autoref{sec:controls}). For the distortion control experiments we test five different conditions, with three fixed $\beta$ variants: (i) $\beta=0$ (MVDR), (ii) $\beta=1$ (MWF) and (iii) $\beta=10$ (Aggressive MWF), and two learned variants: (iv) frequency-dependent and (v) the \ac{SPP}-driven $\beta$ proposed in \autoref{sec:controls}.

Note that the diagram in \autoref{fig:system_overview} shows that $\alpha_{ss}$ and $\alpha_{nn}$ can be estimated online, we found no advantage of doing so in our experiments, and therefore only present experiments where they are learned during training and fixed during inference.

\vspace{0.2cm}
\subsection{Experiment 3: Baseline comparison}
Here we compare the proposed NeuralPMWF method with several low-compute neural network baselines. Firstly, we replicated the multichannel speech-enhancement MC-CRN baseline from \cite{Xu2024}, an extension of \cite{Tan18} that incorporates spatial information using $20$ spatially uniformly distributed maxDI beamformers \cite{Donley2021}. MC-CRN modifies the \ac{ERB} features of the reference channel with a real-gain. Furthermore, we adapted the low-compute GTCRN \cite{Rong2024} to handle multichannel audio by concatenating input channels and increasing the number of input filters. Finally, include the TinyGRU+MWF baseline proposed in \cite{Pandey2024}, which estimates a single-channel complex-valued mask used in conjunction with Wang et. al's \ac{MWF} formulation \cite{Wang2023}.

\begin{table}[h]
\centering
\begin{adjustbox}{width=1\columnwidth}
\begin{tabular}{|c|c|c|c|c|c|c|}
\hline
\textbf{Method} & \textbf{STOI} & \textbf{SI-SDR} & \textbf{SNR} & \textbf{NB-PESQ} & \textbf{MMACs} & \textbf{Params.} \\ \hline
\textbf{NeuralPMWF (ours)} & \textbf{74.3} & \textbf{5.5} & \textbf{6.91} & \textbf{2.12} & 24.95 & 164.9k \\ \hline
TinyGRU+MWF & 73.1 & 5.02 & 6.56 & 2.08 & 24.37 & 152.8k \\ \hline
GTCRN+MWF & 72.7 & 4.66 & 6.23 & 2.01 & 91 & 26.5k \\ \hline
MCCRN+MWF & 69.9 & 3.88 & 5.78 & 1.79 & 35 & 117k \\ \hline
Input & 58.1 & -2.83 & -2.84 & 1.47 & -- & -- \\ \hline
\end{tabular}
\end{adjustbox}
\vspace{0.2cm}
\caption{Comparison of the proposed NeuralPMWF and different baselines. SI-SDR and SNR are expressed in dB. MMACs stands for million multiply-accumulate operations per second.}
\label{tab:baselines}
\end{table}

\begin{table}[h]
\centering
\begin{adjustbox}{width=1\columnwidth}
\begin{tabular}{|c|c|c|c|c|}
\hline
\multicolumn{5}{|c|}{(a) $\alpha$: \textbf{Smoothing Control}} \\ \hline
\textbf{Method} & \textbf{STOI} & \textbf{SI-SDR} & \textbf{SNR} & \textbf{NB-PESQ} \\ \hline
Cum. mean & 69.8 & 3.62 & 5.57 & 1.93 \\ \hline
Fixed & 71.8 & 4.83 & 6.40 & 2.04 \\ \hline
Frequency-dependent & \textbf{72.3} & \textbf{4.90} & \textbf{6.46} & \textbf{2.06} \\ \hline
SPP-driven  & 71.9 & 4.90 & 6.45 & 2.05 \\ \hline
\multicolumn{5}{|c|}{(b) $\beta$: \textbf{Distortion Control}} \\ \hline
$\beta = 0$ (MVDR) & 69.8 & 3.62 & 5.57 & 1.93 \\ \hline
$\beta = 1$ (MWF) & 69.9 & 3.62 & 5.57 & 1.93 \\ \hline
$\beta = 10$ & 67.2 & 1.71 & 4.42 & 1.87 \\ \hline
Frequency-dependent & 70.0 & 3.70 & 5.62 & 1.93 \\ \hline
SPP-driven & \textbf{74.3} & \textbf{5.5} & \textbf{6.91} & \textbf{2.12}  \\ \hline

\end{tabular}
\end{adjustbox}
\vspace{0.2cm}
\caption{(a) smoothing ($\alpha_{nn}$ and $\alpha_{ss}$) and (b) distortion ($\beta$) strategy comparison. In (a), several $\beta$ strategies are tested while cumulative mean covariance smoothing was used. In (b), several $\alpha$ strategies are used while $\beta$ was fixed at $0$ (MVDR)}
\label{tab:ablation}
\end{table}

\begin{figure}
    \centering
    \includegraphics[width=1\columnwidth]{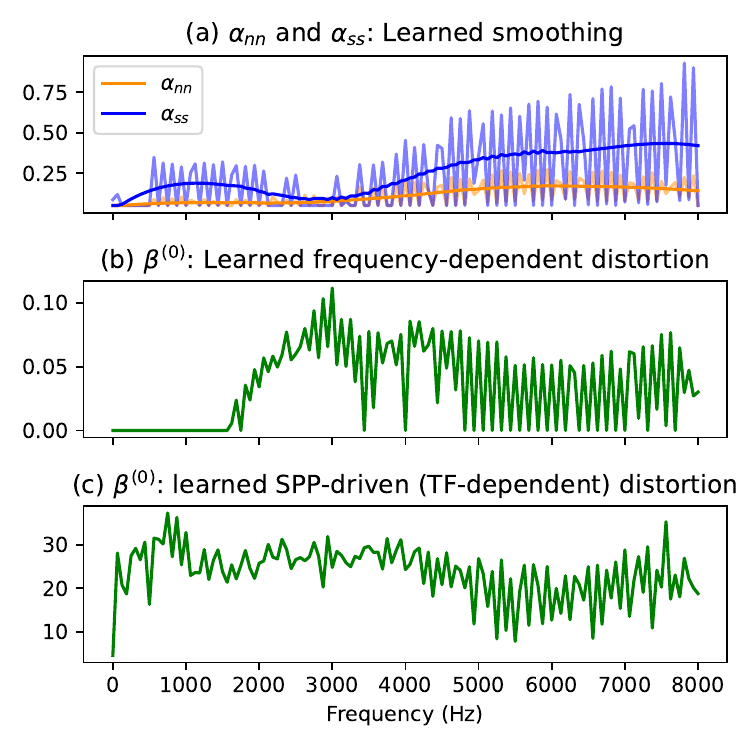}
    \caption{(a): Learned speech and noise covariance smoothing parameters per frequency. The darker curve shows the interpolated filter using a Savitsky-Golay filter to exhibit a trend. (b): $\beta^{(0)}$ results when training using SPP-independent $\beta$. (c): $\beta^{(0)}$ for SPP-dependent $\beta^{(0)}$.}
    \label{fig:alpha-beta}
\end{figure}

\section{Results} \label{sec:results}

The baseline comparison results shown in \autoref{tab:baselines} reveal that NeuralPMWF surpasses all baselines in all metrics, while requiring very low computational resources. Turning to the ablation study results in \autoref{tab:ablation}-a, cumulative mean showed the worst performance among all smoothing techniques, explained by its diminishing ability to adapt to acoustic changes over time. Other exponential smoothing methods performed similarly. NeuralPMWF is also attractive due to its explainability, as the learned internal parameters can be visualized, as shown in \autoref{fig:alpha-beta}. In \autoref{fig:alpha-beta}a, the speech smoothing parameter $\alpha_{ss}$ is bigger than the noise's $\alpha_{nn}$, in accordance with the classical assumption that speech statistics change faster than noise \cite{Cohen2001}.

Turning to the distortion control experiments shown in \autoref{tab:ablation}-b, agressive \ac{MWF} ($\beta=10$) performed the worst among all variants. The \ac{MVDR} and \ac{MWF} showed almost identical results. Finally, expressive performance improvements were obtained by applying SPP-driven $\beta$, demonstrating the importance of dynamic, time-varying distortion control. As shown in \autoref{fig:alpha-beta}-c, very agressive filtering ($\beta > 30$) can be applied if the network is certain of speech absence. \autoref{fig:alpha-beta}-b shows that learning a time-invariant frequency-dependent value for $\beta$ results in low distortion for low frequencies, where most speech energy is typically located \cite{Rabiner1993}.

\section{Conclusion and future steps} \label{sec:conclusion}
We presented NeuralPMWF, a novel system that utilizes a tiny neural network to control the classical beamformer algorithm \ac{PMWF}. Our system is trained end-to-end to estimate noise and speech spatial characteristics, while dynamically controlling the \ac{PMWF}'s output distortion and the covariance smoothing parameters in real-time, resulting in increased speech quality when comparing to competitive low-compute multichannel baselines. Future research directions include using alternative PMWF formulations and further optimizing the network MIMO architecture, as well as investigating the method's generalization to geometries other than the one it was trained on.

\bibliographystyle{IEEEtran}

\bibliography{main}

\begin{thebibliography}{10}
\providecommand{\url}[1]{#1}
\csname url@samestyle\endcsname
\providecommand{\newblock}{\relax}
\providecommand{\bibinfo}[2]{#2}
\providecommand{\BIBentrySTDinterwordspacing}{\spaceskip=0pt\relax}
\providecommand{\BIBentryALTinterwordstretchfactor}{4}
\providecommand{\BIBentryALTinterwordspacing}{\spaceskip=\fontdimen2\font plus
\BIBentryALTinterwordstretchfactor\fontdimen3\font minus \fontdimen4\font\relax}
\providecommand{\BIBforeignlanguage}[2]{{%
\expandafter\ifx\csname l@#1\endcsname\relax
\typeout{** WARNING: IEEEtran.bst: No hyphenation pattern has been}%
\typeout{** loaded for the language `#1'. Using the pattern for}%
\typeout{** the default language instead.}%
\else
\language=\csname l@#1\endcsname
\fi
#2}}
\providecommand{\BIBdecl}{\relax}
\BIBdecl

\bibitem{Haykin2009}
S.~Haykin and K.~J.~R. Liu, \emph{Acoustic Beamforming for Hearing Aid Applications}.\hskip 1em plus 0.5em minus 0.4em\relax Wiley, 2009, pp. 269--302.

\bibitem{Guiraud2022}
P.~Guiraud, S.~Hafezi, P.~A. Naylor, A.~H. Moore, J.~Donley, V.~Tourbabin, and T.~Lunner, ``An introduction to the speech enhancement for augmented reality (spear) challenge,'' in \emph{Proc. Int. Workshop on Acoust. Signal Enhancement ({IWAENC})}, 2022, pp. 1--5.

\bibitem{Souden2010}
M.~Souden, J.~Benesty, and S.~Affes, ``On optimal frequency-domain multichannel linear filtering for noise reduction,'' \emph{IEEE Transactions on Audio, Speech, and Language Processing}, vol.~18, no.~2, pp. 260--276, 2010.

\bibitem{Ephraim1984}
Y.~Ephraim and D.~Malah, ``Speech enhancement using a minimum-mean square error short-time spectral amplitude estimator,'' \emph{{IEEE} Trans. Acoust., Speech, Signal Process.}, vol.~32, no.~6, pp. 1109--1121, 1984.

\bibitem{Benesty2008}
J.~Benesty, \emph{Microphone array signal processing}.\hskip 1em plus 0.5em minus 0.4em\relax Springer Verlag, 2008.

\bibitem{Tashev2009}
I.~J. Tashev, \emph{Sound Capture and Processing: Practical Approaches}.\hskip 1em plus 0.5em minus 0.4em\relax Wiley Publishing, 2009.

\bibitem{Gannot2017}
S.~Gannot, E.~Vincent, S.~Markovich-Golan, and A.~Ozerov, ``A consolidated perspective on multimicrophone speech enhancement and source separation,'' \emph{{IEEE} Trans. Audio, Speech, Language Process.}, vol.~25, no.~4, pp. 692--730, 2017.

\bibitem{VanTrees2002}
H.~L. Van~Trees, \emph{Optimum array processing: Part IV of detection, estimation, and modulation theory}.\hskip 1em plus 0.5em minus 0.4em\relax John Wiley \& Sons, 2002.

\bibitem{Wang2023}
Z.-Q. Wang, G.~Wichern, S.~Watanabe, and J.~Le~Roux, ``Stft-domain neural speech enhancement with very low algorithmic latency,'' \emph{{IEEE} Trans. Audio, Speech, Language Process.}, vol.~31, pp. 397--410, 2023.

\bibitem{Capon1969}
J.~Capon, ``High-resolution frequency-wavenumber spectrum analysis,'' \emph{Proceedings of the IEEE}, vol.~57, no.~8, pp. 1408--1418, 1969.

\bibitem{Griffiths1982}
L.~Griffiths and C.~Jim, ``An alternative approach to linearly constrained adaptive beamforming,'' \emph{IEEE Transactions on Antennas and Propagation}, vol.~30, no.~1, pp. 27--34, 1982.

\bibitem{Lorenz2005}
R.~Lorenz and S.~Boyd, ``Robust minimum variance beamforming,'' \emph{{IEEE} Trans. Signal Process.}, vol.~53, no.~5, pp. 1684--1696, 2005.

\bibitem{Buckley1986}
K.~Buckley and L.~Griffiths, ``An adaptive generalized sidelobe canceller with derivative constraints,'' \emph{{IEEE} Trans. Antennas Propag.}, vol.~34, no.~3, pp. 311--319, 1986.

\bibitem{Doclo2007}
S.~Doclo, A.~Spriet, J.~Wouters, and M.~Moonen, ``Frequency-domain criterion for the speech distortion weighted multichannel wiener filter for robust noise reduction,'' \emph{Speech Communication}, vol.~49, no. 7-8, pp. 636--656, 2007.

\bibitem{Braun2015}
S.~Braun, K.~Kowalczyk, and E.~A.~P. Habets, ``Residual noise control using a parametric multichannel wiener filter,'' in \emph{2015 IEEE International Conference on Acoustics, Speech and Signal Processing (ICASSP)}, 2015, pp. 360--364.

\bibitem{Bagheri2019}
S.~Bagheri and D.~Giacobello, ``Exploiting multi-channel speech presence probability in parametric multi-channel wiener filter.'' in \emph{Proc. Conf. of Int. Speech Commun. Assoc. ({INTERSPEECH})}, 2019, pp. 101--105.

\bibitem{Ngo2009}
\BIBentryALTinterwordspacing
K.~Ngo, A.~Spriet, M.~Moonen, J.~Wouters, and S.~H. Jensen, ``Incorporating the conditional speech presence probability in multi-channel wiener filter based noise reduction in hearing aids,'' \emph{EURASIP Journal on Advances in Signal Processing}, vol. 2009, p. 930625, 2009. [Online]. Available: \url{https://asp-eurasipjournals.springeropen.com/articles/10.1155/2009/930625}
\BIBentrySTDinterwordspacing

\bibitem{Wang2018}
Z.-Q. Wang, J.~Le~Roux, and J.~R. Hershey, ``Multi-channel deep clustering: Discriminative spectral and spatial embeddings for speaker-independent speech separation,'' in \emph{Proc. {IEEE} Int. Conf. on Acoust., Speech and Signal Process. ({ICASSP})}, 2018, pp. 1--5.

\bibitem{Tolooshams2020}
B.~Tolooshams, R.~Giri, A.~H. Song, U.~Isik, and A.~Krishnaswamy, ``Channel-attention dense u-net for multichannel speech enhancement,'' in \emph{Proc. {IEEE} Int. Conf. on Acoust., Speech and Signal Process. ({ICASSP})}, 2020, pp. 836--840.

\bibitem{Liu2020}
C.-L. Liu, S.-W. Fu, Y.-J. Li, J.-W. Huang, H.-M. Wang, and Y.~Tsao, ``Multichannel speech enhancement by raw waveform-mapping using fully convolutional networks,'' \emph{{IEEE} Trans. Audio, Speech, Language Process.}, vol.~28, pp. 1888--1900, 2020.

\bibitem{Lee2023}
D.~Lee and J.-W. Choi, ``Deft-an: Dense frequency-time attentive network for multichannel speech enhancement,'' \emph{{IEEE} Signal Process. Lett.}, vol.~30, pp. 155--159, 2023.

\bibitem{Hsieh2024}
T.-A. Hsieh, J.~Donley, D.~Wong, B.~Xu, and A.~Pandey, ``On the importance of neural wiener filter for resource efficient multichannel speech enhancement,'' in \emph{Proc. {IEEE} Int. Conf. on Acoust., Speech and Signal Process. ({ICASSP})}, 2024, pp. 12\,181--12\,185.

\bibitem{Tao2024}
S.~Tao, P.~Mowlaee, J.~R. Jensen, and M.~Græsbøll~Christensen, ``Learning-based multi-channel speech presence probability estimation using a low-parameter model and integration with mvdr beamforming for multi-channel speech enhancement,'' in \emph{Proc. Int. Workshop on Acoust. Signal Enhancement ({IWAENC})}, 2024, pp. 100--104.

\bibitem{Pandey2022}
A.~Pandey, B.~Xu, A.~Kumar, J.~Donley, P.~Calamia, and D.~Wang, ``Time-domain ad-hoc array speech enhancement using a triple-path network,'' in \emph{Proc. Conf. of Int. Speech Commun. Assoc. ({INTERSPEECH})}, 2022, pp. 729--733.

\bibitem{Tammen2023}
M.~Tammen and S.~Doclo, ``Parameter estimation procedures for deep multi-frame mvdr filtering for single-microphone speech enhancement,'' \emph{IEEE/ACM Transactions on Audio, Speech, and Language Processing}, vol.~31, pp. 3237--3248, 2023.

\bibitem{Souden2010b}
M.~Souden, J.~Chen, J.~Benesty, and S.~Affes, ``Gaussian model-based multichannel speech presence probability,'' \emph{IEEE Transactions on Audio, Speech, and Language Processing}, vol.~18, no.~5, pp. 1072--1077, 2010.

\bibitem{Cohen2001}
I.~Cohen and B.~Berdugo, ``Speech enhancement for non-stationary noise environments,'' \emph{Signal Processing}, vol.~81, no.~11, pp. 2403--2418, 2001.

\bibitem{Cohen2003}
I.~Cohen, ``Noise spectrum estimation in adverse environments: improved minima controlled recursive averaging,'' \emph{IEEE Transactions on Speech and Audio Processing}, vol.~11, no.~5, pp. 466--475, 2003.

\bibitem{Pandey2024}
A.~Pandey and B.~Xu, ``Decoupled spatial and temporal processing for resource efficient multichannel speech enhancement,'' in \emph{Proc. {IEEE} Int. Conf. on Acoust., Speech and Signal Process. ({ICASSP})}, vol.~2, 2024.

\bibitem{Pandey2024a}
A.~Pandey and J.~Azcarreta, ``Ultra low-compute complex spectral masking for multichannel speech enhancement,'' 2024.

\bibitem{tan2019learning}
K.~Tan and D.~L. Wang, ``Learning complex spectral mapping with gated convolutional recurrent networks for monaural speech enhancement,'' \emph{IEEE/ACM Transactions on Audio, Speech, and Language Processing}, vol.~28, pp. 380--390, 2019.

\bibitem{Chung2014}
J.~Chung, C.~Gulcehre, K.~Cho, and Y.~Bengio, ``Empirical evaluation of gated recurrent neural networks on sequence modeling,'' 2014.

\bibitem{Paszke2019}
A.~Paszke, S.~Gross, F.~Massa, A.~Lerer, J.~Bradbury, G.~Chanan, T.~Killeen, Z.~Lin, N.~Gimelshein, L.~Antiga, A.~Desmaison, A.~Kopf, E.~Yang, Z.~DeVito, M.~Raison, A.~Tejani, S.~Chilamkurthy, B.~Steiner, L.~Fang, J.~Bai, and S.~Chintala, ``Pytorch: An imperative style, high-performance deep learning library,'' in \emph{Proc. Neural Inform. Process. Conf}, H.~Wallach, H.~Larochelle, A.~Beygelzimer, F.~d\textquotesingle Alch\'{e}-Buc, E.~Fox, and R.~Garnett, Eds., vol.~32.\hskip 1em plus 0.5em minus 0.4em\relax Curran Associates, Inc., 2019.

\bibitem{Pandey2021}
A.~Pandey and D.~Wang, ``Dense cnn with self-attention for time-domain speech enhancement,'' \emph{{IEEE} Trans. Audio, Speech, Language Process.}, vol.~29, pp. 1270--1279, 2021.

\bibitem{Kingma2014}
D.~P. Kingma, ``Adam: A method for stochastic optimization,'' in \emph{Proc. Int. Conf. on Learning Representations}, 2014.

\bibitem{Reddi2018}
S.~J. Reddi, S.~Kale, and S.~Kumar, ``On the convergence of adam and beyond,'' in \emph{Proc. Int. Conf. on Learning Representations}, 2018.

\bibitem{Taal2011}
C.~H. Taal, R.~C. Hendriks, R.~Heusdens, and J.~Jensen, ``An algorithm for intelligibility prediction of time–frequency weighted noisy speech,'' \emph{IEEE Transactions on Audio, Speech, and Language Processing}, vol.~19, no.~7, pp. 2125--2136, 2011.

\bibitem{Rix2001}
A.~Rix, J.~Beerends, M.~Hollier, and A.~Hekstra, ``Perceptual evaluation of speech quality (pesq)-a new method for speech quality assessment of telephone networks and codecs,'' in \emph{Proc. {IEEE} Int. Conf. on Acoust., Speech and Signal Process. ({ICASSP})}, vol.~2, 2001, pp. 749--752 vol.2.

\bibitem{Vincent2006}
E.~Vincent, R.~Gribonval, and C.~Fevotte, ``Performance measurement in blind audio source separation,'' \emph{{IEEE} Trans. Audio, Speech, Language Process.}, vol.~14, no.~4, pp. 1462--1469, 2006.

\bibitem{Reddy2020}
C.~K. Reddy, V.~Gopal, R.~Cutler, E.~Beyrami, R.~Cheng, H.~Dubey, S.~Matusevych, R.~Aichner, A.~Aazami, S.~Braun \emph{et~al.}, ``The interspeech 2020 deep noise suppression challenge: Datasets, subjective testing framework, and challenge results,'' in \emph{Proc. Conf. of Int. Speech Commun. Assoc. ({INTERSPEECH})}, 2020.

\bibitem{Allen1979}
J.~B. Allen and D.~A. Berkley, ``{Image method for efficiently simulating small‐room acoustics},'' \emph{J. Acoust. Soc. Am.}, vol.~65, no.~4, pp. 943--950, 04 1979.

\bibitem{Scheibler2018}
R.~Scheibler, E.~Bezzam, and I.~Dokmanić, ``Pyroomacoustics: A python package for audio room simulation and array processing algorithms,'' in \emph{Proc. {IEEE} Int. Conf. on Acoust., Speech and Signal Process. ({ICASSP})}, 2018, pp. 351--355.

\bibitem{Xu2024}
Z.~Xu, A.~Aroudi, K.~Tan, A.~Pandey, J.-S. Lee, B.~Xu, and F.~Nesta, ``Fovnet: Configurable field-of-view speech enhancement with low computation and distortion for smart glasses,'' in \emph{Proc. Conf. of Int. Speech Commun. Assoc. ({INTERSPEECH})}, 2024, pp. 3350--3354.

\bibitem{Tan18}
K.~Tan and D.~Wang, ``A convolutional recurrent neural network for real-time speech enhancement,'' in \emph{Proc. Conf. of Int. Speech Commun. Assoc. ({INTERSPEECH})}, 2018, pp. 3229--3233.

\bibitem{Donley2021}
J.~Donley, V.~Tourbabin, J.-S. Lee, M.~Broyles, H.~Jiang, J.~Shen, M.~Pantic, V.~K. Ithapu, and R.~Mehra, ``Easycom: An augmented reality dataset to support algorithms for easy communication in noisy environments,'' 2021.

\bibitem{Rong2024}
X.~Rong, T.~Sun, X.~Zhang, Y.~Hu, C.~Zhu, and J.~Lu, ``Gtcrn: A speech enhancement model requiring ultralow computational resources,'' in \emph{Proc. {IEEE} Int. Conf. on Acoust., Speech and Signal Process. ({ICASSP})}, 2024, pp. 971--975.

\bibitem{Rabiner1993}
L.~Rabiner and B.-H. Juang, \emph{Fundamentals of speech recognition}.\hskip 1em plus 0.5em minus 0.4em\relax Prentice-Hall, Inc., 1993.

\end{thebibliography}

\end{document}